\documentclass{aastex63}[RNAAS]

\usepackage{soul}
\usepackage[colorinlistoftodos,textwidth=14mm,textsize=scriptsize]{todonotes}

\shorttitle{AGN variability observables in LSST}
\shortauthors{Kova{\v c}evi{\'c} et al.}
\graphicspath{{./}{figures/}}
\begin{document}

\title{LSST AGN SC Cadence Note: Two metrics on AGN variability observables}

\author{Andjelka Kova{\v c}evi{\'c}, Dragana Ili{\'c}, Isidora Jankov, Luka {\v C}. Popovi{\'c}, Ilsang Yoon,}
\author{Viktor Radovi{\'c}, Neven Caplar , Iva {\v C}vorovi{\'c}-Hajdinjak -- on behalf of the AGN SC}

\section{Executive Summary}

We have developed {two metrics} related to AGN variability observables (time-lags, periodicity, and Structure Function (SF)) to evaluate { LSST} OpSim FBS 1.5, 1.6, 1.7 performance in AGN time-domain analysis. For this purpose, we generate an ensemble of {AGN }light curves based on AGN empirical relations and { LSST} OpSim cadences.
Although our metrics show that denser LSST cadences produce more reliable time-lag, periodicity, and SF measurements, the discrepancies in the performance between different LSST OpSim cadences are not drastic based on Kullback-Leibler divergence.
This is complementary to Yu and Richards results {on DCR and SF metrics}  (see Yu's talk \url{https://docs.google.com/presentation/d/12Q1zKiWtoQAXsh7GS6J9TYhEKsWfkN5sVDufEyCKkEE/edit#slide=id.gc6954dd1ce_0_3}), 
extending them to include the point of view of AGN variability.
%Using machine learning methods for time-lag and periodicity measurements, we also demonstrate that poll of lower density   OpSim cadences can serve as a test-bed for  (deep) machine learning.

\section{The Two Metrics} \label{sec:intro}

A detailed description of metric analysis performed on OpSim 1.5 and 1.6 is given in Kova{\v c}evi{\'c} et al. {(submitted version can be found in} \url{https://github.com/LSST-sersag/white_paper/blob/main/data/paper.pdf}).  Results based on the OpSim 1.7 are available at  \url{https://github.com/LSST-sersag/agn_cadences}.

\subsection{Metrics based on AGN time lag (\texorpdfstring{$\tau$}{tau}) and periodicities (P) measurements}\label{lagmetr}

%The measured physical parameters of the AGN ensemble should be concentrated in a subspace of the space of all possible (in a statistical sense) AGN data.
We designed a metric $\log \phi_{\mathcal{T}}$ for time-lags and periodicities that identifies smooth trends (i.e. hyperplanes or linear relations) in complex datasets. This scaling relation ties relative formal error of measured time scale   $\mathcal{T}_\mathrm{obs}\in(\tau, P)$ to {the characteristics of AGN light curve and OpSim cadendes}:
  the amplitude of flux variation $F_\mathrm{var}$ relative to the measured flux (or photometric) error $\sigma$:  ($\frac{F_\mathrm{var}}{\sigma}$); the ratio of observed time scale $\mathcal{T}_\mathrm{obs}\in(\tau, P)$  and light curve sampling time $\Delta t$:  $\frac{\mathcal{T}_\mathrm{obs}}{\Delta t}$. 
The proposed metric is:
\begin{equation}
\log \phi_{\mathcal{T}}=\log\frac{\sigma_{\mathcal{T}}}{\mathcal{T}}\propto A+C_{1}\frac{F_\mathrm{var}}{\sigma}+C_{2}\frac{ \mathcal{T}_\mathrm{obs}}{(1+z)\Delta t},
\label{89}
\end{equation}
\noindent where we assume that the error of {the rest-frame} time lag or periodicity inferred from the observed light curve will be increasing with an increasing redshift of the object for the same observed time scale and sampling time. 
The metric coefficients tell us how to {compare variables}  on a given ensemble of objects.
Measuring the observables and their uncertainties for each object, we can estimate metric (i.e., coefficients $A, C_{1}, C_{2}$)  for given AGN light curves ensemble generated from OpSim cadences and {probability density function (PDF) of the metric}.

\subsection{Structure function  metric}

The idea of a  metric is to estimate deviations of SFs based on OpSim cadences (termed "gappy" light curve, $SF_{\rm gappy}$ from 'reference' SF obtained from  densely and uniformly sampled light curve ($SF_{\rm conti}$).
Let  {an ensemble of $k$ simulated}  light curves be generated, {providing} $SF^{i}_{\rm conti} $and $SF^{i}_{\rm gappy}, i=1, k$ where k is the number of simulated light curves. Then simplistic {metric} of similarity between $SF^{i}_{\rm conti} $ and $SF^{i}_{\rm gappy}, i=1, k$ curves is given by
$ M^{i}=(SF^{i}_{\rm conti} -SF^{i}_{\rm gappy})^{2}$,
\noindent where $M^{i}$ is the {ensemble} of the deviations {curves of SF}.
However, we can generalize this concept. 
We calculate these deviations curves for objects in  different redshift bins and  average the deviations curves for each redshift bin $z$:
 \begin{equation}
 M^{2}_{z}=\frac{1}{N_z}\sum (SF^{i}_{\rm conti} -SF^{i}_{\rm gappy})^{2}=\frac{1}{N_z}\sum^{N_{z}}_{i=1} M^{2}_{i},
\label{eq:66}
\end{equation}
\noindent where {$M^{2}_{z}$ is an averaged deviation curve for redshift bin $z$},  $N_z$ is the number of deviations curves $M^{i}$ within the redshift bin z.
We can plot thesse {averaged SF deviations} curves  {on the redshift and characteristics time scale domain} as 2D maps or to calculate PDFs on both domains.

\section{Analysis of Results}
We tested our metrics on simulated AGN light curves. More details in Kova\v cevi\'c et al. and on the GitHub (\url{https://github.com/LSST-sersag/white_paper/blob/main/data/paper.pdf})
\subsection{Metrics based on time lags and periodicities}
 We  {learned} the time lags and periodicities metrics and estimated probability density (PDF) of performances of these metrics for simulated objects based on g and r-band  LSST OpSim cadences  (see auxiliary Table 1 in  \url{https://github.com/LSST-sersag/white_paper/blob/main/data/table_1.pdf}). 
 How much given distributions are away from each other we estimated using Kullback-Leibler (KL) Divergence $D_{KL}(pdf1||pdf2)$.
 
\begin{figure*}
\gridline{\fig{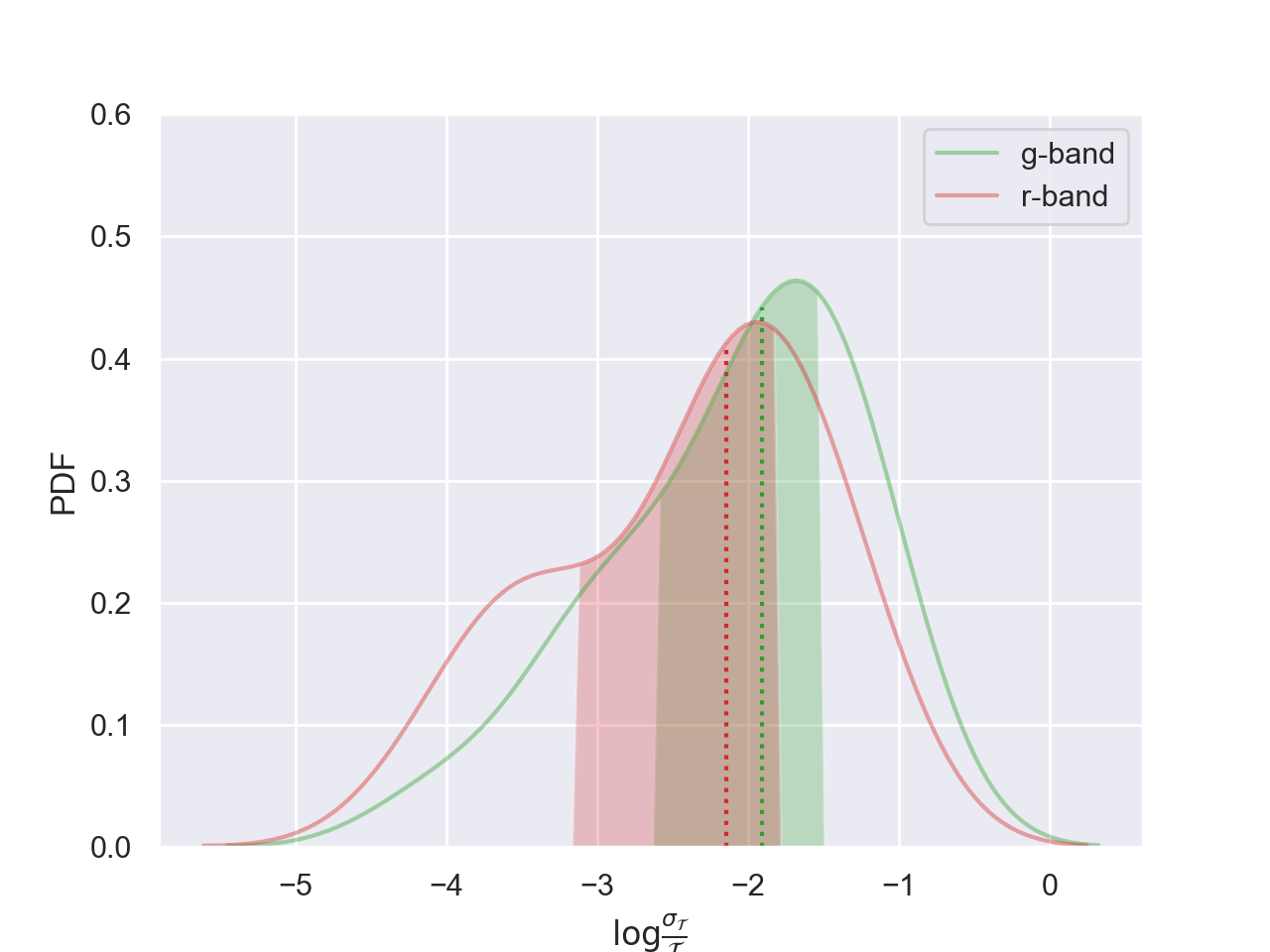}{0.43\textwidth}{\vspace{-1.5em}(a)}
          \fig{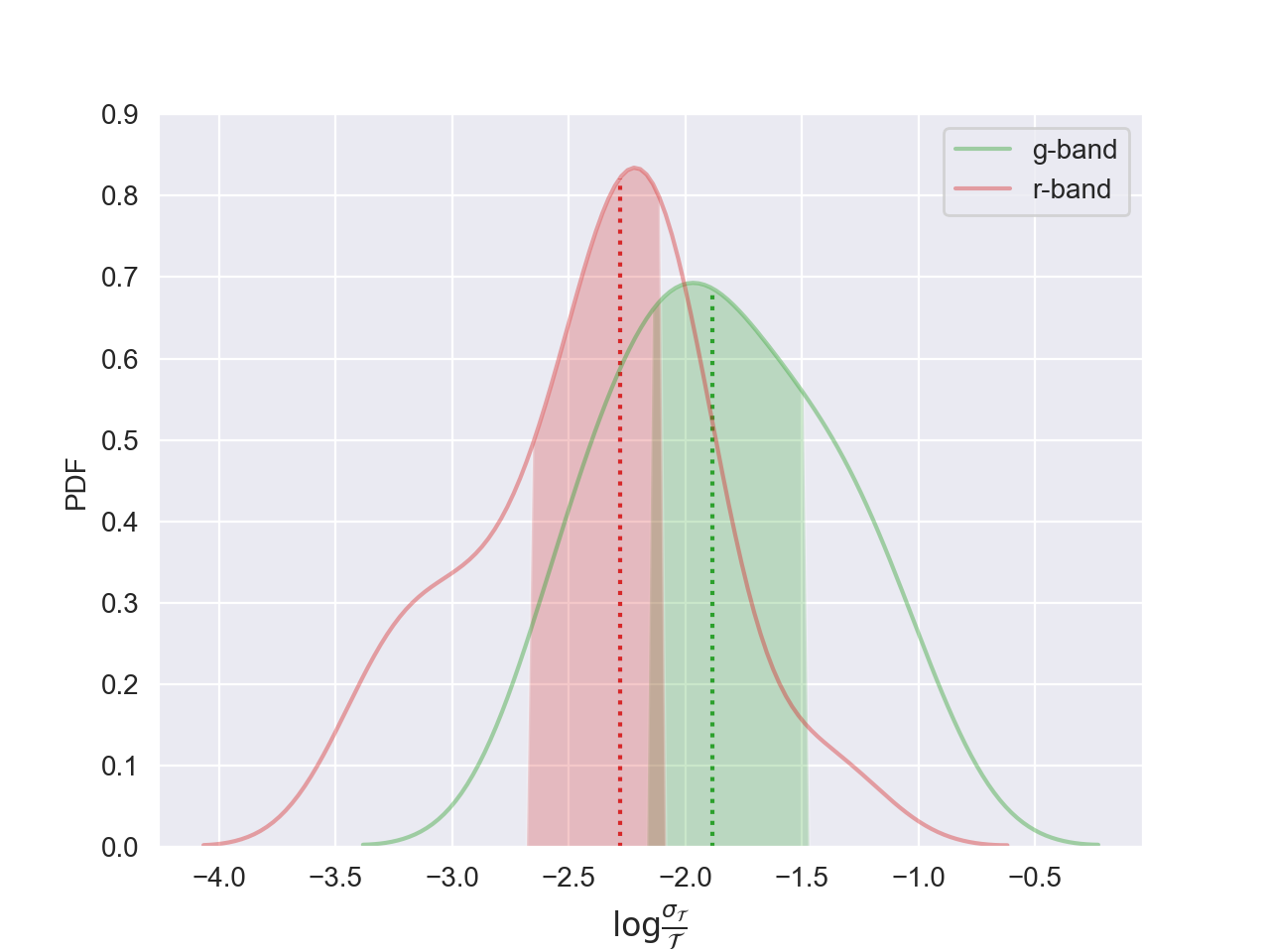}{0.43\textwidth}{\vspace{-1.5em}(b)}
          }
          \vspace{-1.5em}
\caption{\small Densities of metrics for the time lags (a) and periodicities (b) measured on the artificial set of objects (see an auxiliary Table 1 at  \url{https://github.com/LSST-sersag/white_paper}). Vertical dashed lines denote the medians of distributions, and shaded regions correspond to ranges of 25 and 75 percentiles. 
\label{fig:lagspers}}
\vspace{-1.5em}
\end{figure*}

 Figure \ref{fig:lagspers}(a) displays the PDF of time lag metrics for g and r-band objects.
 Here the KL divergence is almost symmetric  $D^{\tau}_{KL}(pdf_g||pdf_r)\sim D^{\tau}_{KL}(pdf_r||pdf_g)\sim0.02$
indicating that our metric detected a slight difference of time-lag measurement in g and r-band. A simple visual inspection of the plot indicates that the smaller relative formal errors  (and their median) are more likely to occur for the r- band light curves.
 However, metrics for periodicities measurement (see Fig. \ref{fig:lagspers}(b))  are more sensitive to sampling density as KL- divergence is two times larger  $D^{P}_{KL}(pdf_g||pdf_r)=0.04$. The medians of metric performance in two bands are at a larger mutual distance than those for time lags. Similar to time lags,   more reliable oscillation detection is expected in the r-band. This is not surprising since the r-band contains almost three times denser cadences, enabling light curves to capture more information from underlying oscillations. The  DDF cadences are very likely to produce confident time-lag and periodicity detections   (see Kova{\v c}evi{\'c} et al).

\subsection{ Structure Function Metric}

\begin{figure*}
\gridline{\fig{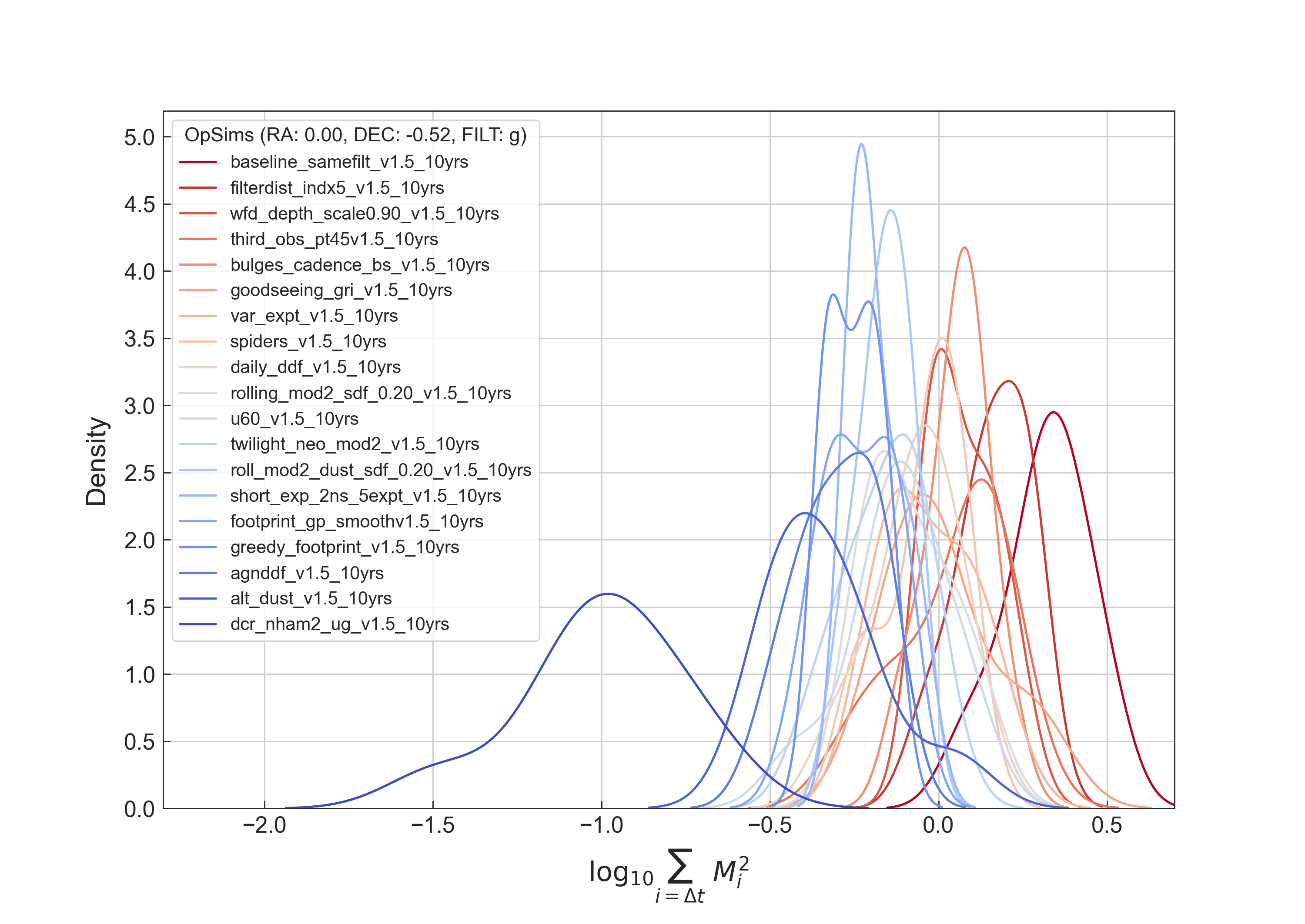}{0.43\textwidth}{\vspace{-1.5em}(a)}
          \fig{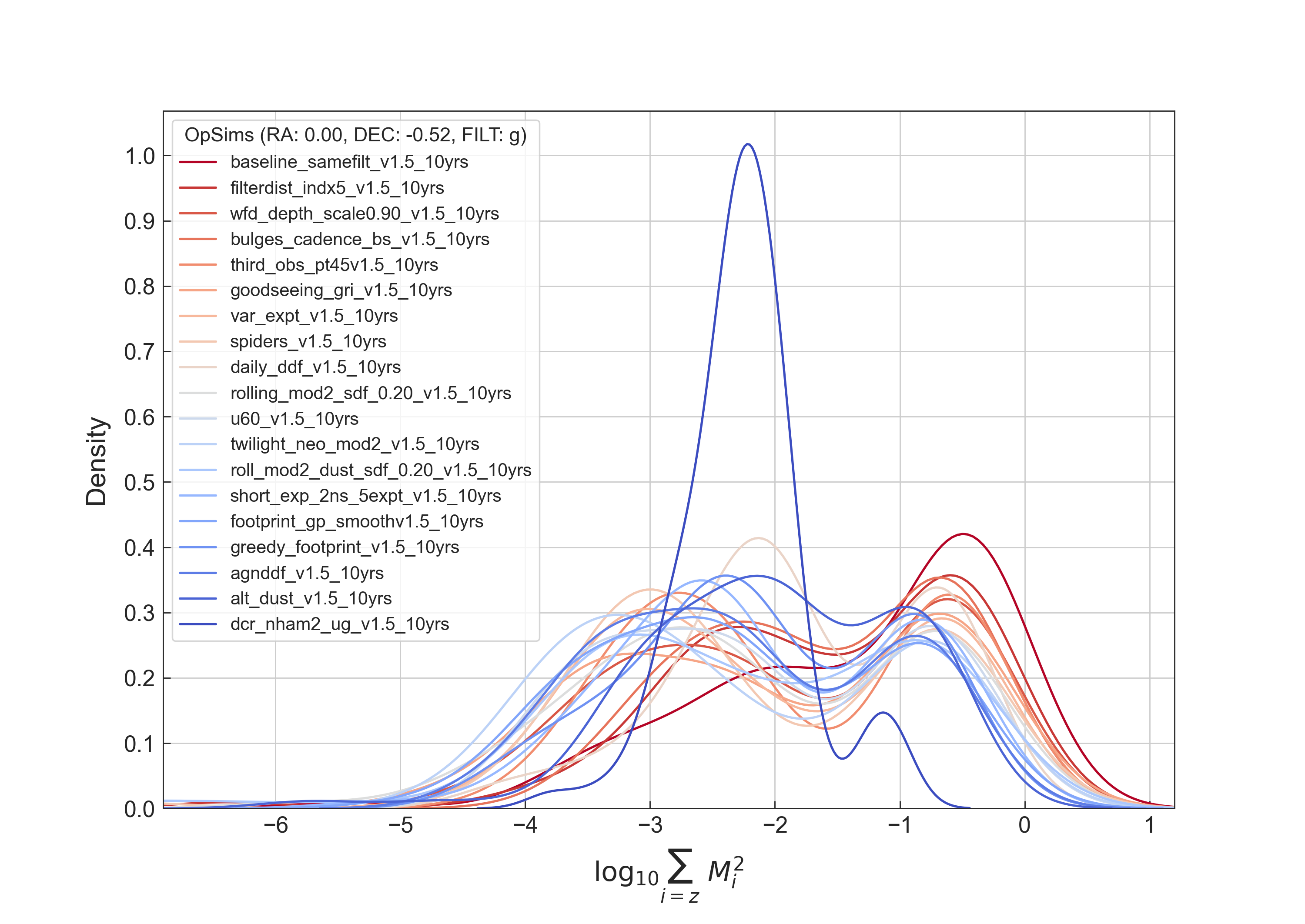}{0.43\textwidth}{\vspace{-1.5em}(b)}
          }
          \vspace{-2.0em}
\gridline{\fig{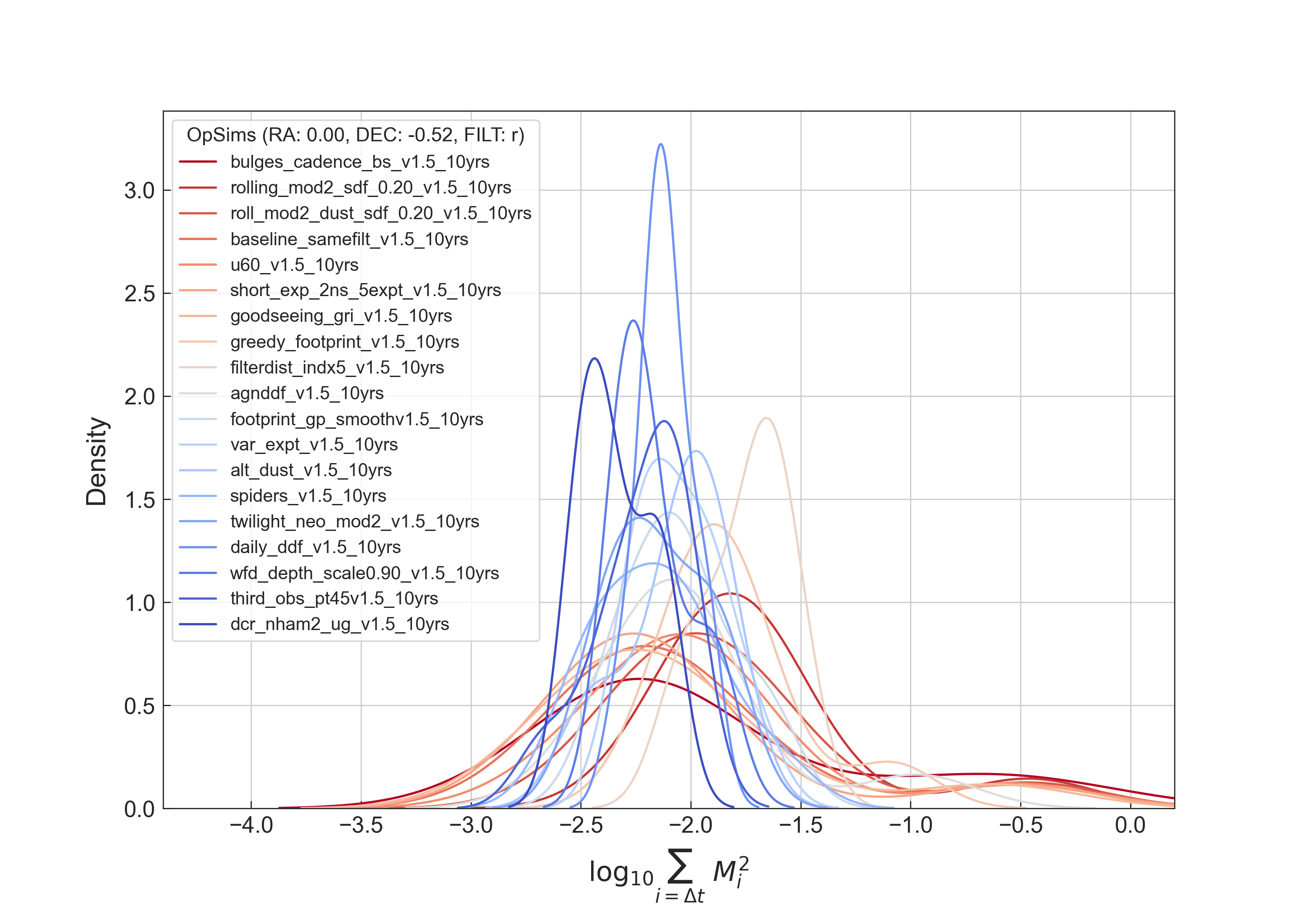}{0.43\textwidth}{\vspace{-1.5em}(c)}
          \fig{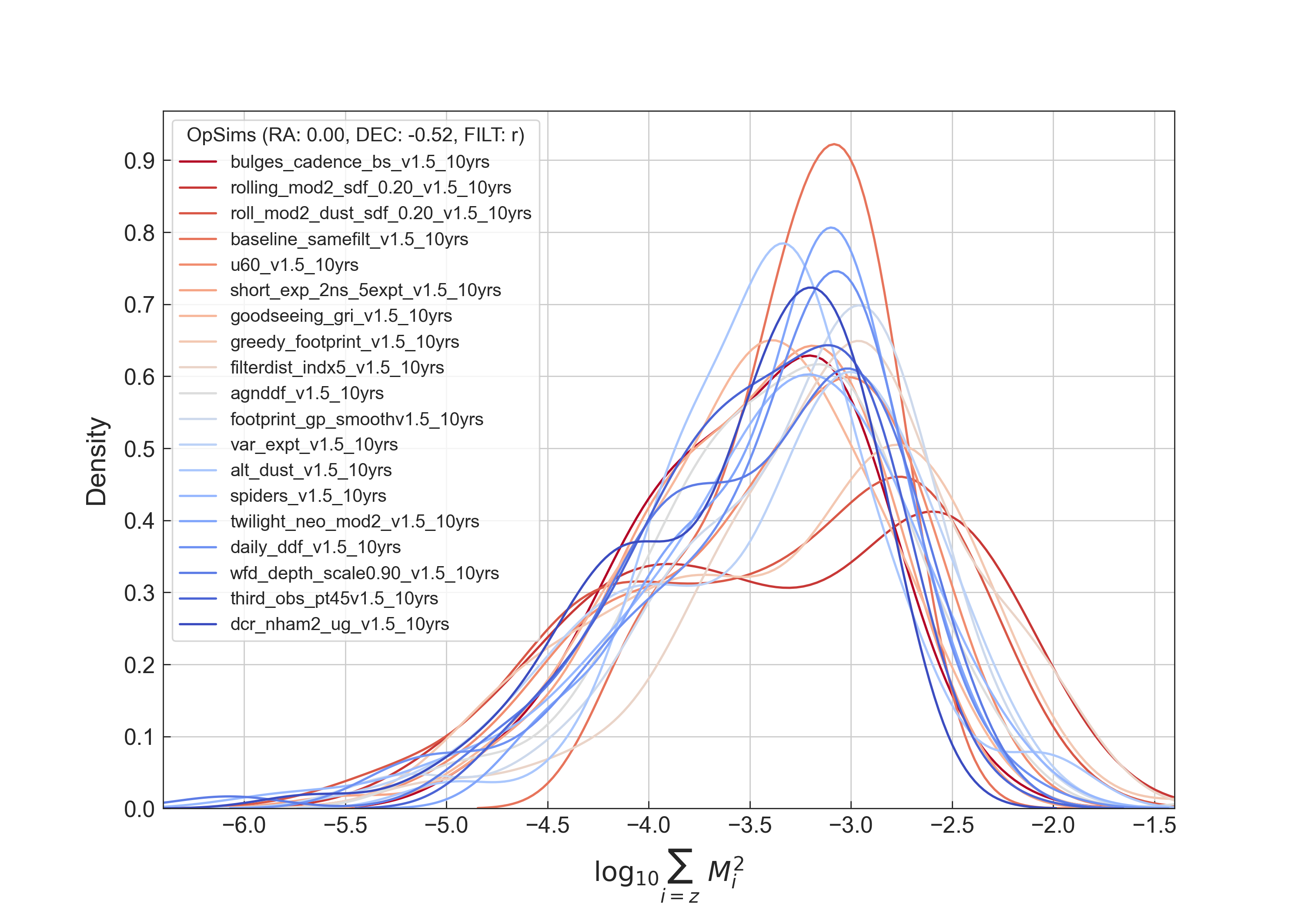}{0.43\textwidth}{\vspace{-1.5em}(d)}
          }
          \vspace{-1.5em}
\caption{\small Densities of summed SF-metric in g-band (upper row) and r-band (lower row): (a),(c) across SF-time scale ($\Delta t$) and (b), (d) redshifts. OpSim cadences are given in legends. Cooler colors correspond to the OpSim cadence  with the smaller metric mean value 
(least deviation), while the warmer colors denote a larger metric mean value.  }
\label{fig:sfmw}
\vspace{-1.5em}
\end{figure*}

Figure \ref{fig:sfmw}(a)-(d) present PDFs of SF-metric for each { LSST} OpSim cadence summed across
the SF time scales $\Delta t$ (left column of plots) and redshifts (right column of plots). The peak of each distribution shows the most likely metric value, i.e., square of deviation of SF based on { LSST} OpSim cadence from 'reference' SF based on ideal cadence. PDFs are marked with different colors for each { LSST} OpSim cadence, so that bluer corresponds to the { LSST} OpSim cadence  with the smallest mean metric value (least deviation), while  the redder color 
distinguishes the { LSST} OpSims cadences that have the immense mean metric  value (highest deviations). The metric in r-band shows that SFs are more coherent with more minor deviations than those in g-band along the SF-time scale ($\Delta t$, compare plots (a) vs. (c)). Similar behavior can be seen across objects' redshifts (compare (b) vs. (d)).
KL-divergence of SF metric in g vs. r-band is given in auxiliary Table 2 (\url{https://github.com/LSST-sersag/white_paper/blob/main/data/table_2.pdf}).  Cadences, baseline\_samefilt\_v1.5\_10yrs
and bulges\_cadence\_bs\_v1.5\_10yrs produce significant KL-divergence of metrics in g and r-band across redshifts.  The  DDF cadences are very likely to produce more reliable SFs  (see Kova{\v c}evi{\'c} et al).
{The mean sampling of  u-band cadences, such as $baseline\_u\_ra\_0.0\_de\_-10.0,baseline\_u\_ra\_0.0\_de\_-30.0, rolling\_cadence\_0.8\_u\_ra\_0.0\_de\_-10.0., rolling\_cadence\_0.8\_u\_ra\_0.0\_de\_-30.0$, are 58.2, 75.1, 64.8, 66.7 days respectively. Such condition limit the interpretation of the AGN time-lag and periodicity analysis.}

%More specifically, we determine distributions of individual %SF-metrics along SF-time scales and redshift  for each OpSim %cadence strategy (see Figure \ref{fig:sfmw2}). This gives  %insight to individual behavior of SF-metric along given vectors %for each OpSim cadence, i.e. to locate the density of  %deviations in given time scales and redshifts.
%The densities of individual metric on each OpSim cadence across %both vectors (time scales and redshift) are more coherent in %g-band. 

%This result  is not in contradiction with our previous %observation that a summed SF- metric values across time scales %and redshifts are more coherent in r-band, because g-band light %curves cadences have more coherent gaps distributions and  %three times smaller sampling than  r-bands light curves.

%\begin{figure*}
%\gridline{\fig{Dist_g_dt.png}{0.48\textwidth}{(a)}
%          \fig{Dist_g_z.png}{0.48\textwidth}{(b)}
%          }
%\gridline{\fig{Dist_r_dt.png}{0.48\textwidth}{(c)}
%          \fig{Dist_r_z.png}{0.48\textwidth}{(d)}
 %         }
%\caption{}
%\label{fig:sfmw2}
%\end{figure*}

\section{Answers to Questions}
%\iyinline{Have you tried this analysis for %u-band light curve? Since u-band observing %cadences are significantly different for %different OpSims, it might be worth to check if %these metrics pick up any particular set of %OpSims that significantly out- and %under-perform.}
\begin{quote}
{\footnotesize Q1: Are there any science drivers that would strongly argue for, or against, increasing the WFD footprint from 18,000 sq. deg. to 20,000 sq.deg.? Note that the resulting number of visits per pointing would drop by about 10\%. If available, please mention specific simulated cadences, and specific metrics, that support your answer.}
\end{quote}
\vspace{-0.35cm}
{ Answer 1 Not applicable}
\vspace{-0.35cm}
\begin{quote}
{\footnotesize Q2: Assuming that current system performance estimates will hold up, we plan to utilize the additional observing time (which may be as much as 10\% of the survey observing time) for visits for the mini-surveys and the DDFs (with an implicit assumption that the main WFD survey meeting SRD requirements will always be the first priority). What is the best scientific use of this time? If available, please mention specific simulated cadences, and specific metrics, that support your answer.}
\end{quote}
\vspace{-0.35cm}
{ Answer 2 Our detailed analysis in Kova{\v c}evi{\'c} et al suggest that DDF cadences are already very promising for determination of the reliable sizes of region sourounding supermassive black holes (and implicitly masses of these objects), detection of close-binary supermassive black holes candidates and reliable reconstruction of SF.
In the long run, AGN variablity studies constrainig AGN models and revealing nano-Hz gravitational wave candidates-will benefit if these DDF visits are enhanced rather than suppressed.}
\vspace{-0.35cm}
\begin{quote}
{\footnotesize Q3:  Are there any science drivers that would strongly argue for, or against, the  proposal to change the u band exposure from 2x15 sec to 1x50 sec? If available, please mention specific simulated cadences, and specific metrics, that support your answer.} 
\end{quote}
\vspace{-0.35cm}
{ Answer 3 Not applicable.}
\begin{quote}
\vspace{-0.35cm}
{\footnotesize Q4:  Are there any science drivers that would strongly argue for, or against, further changes in observing time allocation per band (e.g., skewed much more towards the blue or the red side of the spectrum)? If available, please mention specific simulated cadences, and specific metrics, that support your answer.}
\end{quote}
\vspace{-0.35cm}
{ Answer 4 For  size-determinations  of regions surrounding AGN (and their masses consequently) our metric assumes that filters can catch both continuum and emission lines. However, different filters will cover different distances from continuum sources. In that sense we anticipated non-skewed time allocation per band. }
\vspace{-0.35cm}
\begin{quote}
{\footnotesize Q5:  Are there any science drivers that would strongly argue for, or against, obtained two visits in a pair in the same (or different) filter? Or the benefits or drawbacks of dedicating a portion of each night to obtaining a third (triplet) visit?  If available, please mention specific simulated cadences, and specific metrics, that support your answer.}
\end{quote}
\vspace{-0.35cm}
 Answer 5 { Photometric reverberation-mapping is based on simultaneous monitoring in different filters since different emission lines contribute at different level \href{https://ui.adsabs.harvard.edu/abs/2012ApJ...747...62C/abstract}{(Chelouche \& Daniel 2012)}, and our proposed metric probes this principle (see discussion in Kova\v cevi\'c et al. 2021).
For example, in some photometric campaigns, measured flux through r band  covers continuum + H$\alpha$ line, and r band covers continuum plus H$\beta$ line. Line  contributions were at several percent, while i-band had contributions from the continuum alone.  Based on the literature, this would be similar for objects with low redshifts. For objects at higher redshift where Balmer lines continually
shift to longer wavelengths, filters must be carefully applied. Our proposed metric probes this principle. Having this in mind, visits in a pair (or in triplet) in different filters will be of particular advantage.
Thus visits in a pair (or in triplet) in different filters will be of particular advantage.}
\vspace{-0.35cm}
\begin{quote}
{\footnotesize Q6:  Are there any science drivers that would strongly argue for, or against, the rolling cadence scenario? Or for or against varying the season length? Or for or against the AltSched N/S nightly pattern of visits? If available, please mention specific simulated cadences, and specific metrics, that support your answer.}
\end{quote}
\vspace{-0.35cm}
{Answer 6
We simulated AGN variability metric for rolling cadence scenario in Kova{\v c}evi{\'c} et al.
From  the OpSim 1.6 we used rolling cadences. Results based on the newest cadence release, OpSim 1.7, are given at  (\url{https://github.com/LSST-sersag/agn_cadences}).
 For example, OpSim\_ roll.cad\_0.8\_g\_RA\_0.0\_D\_-10.0   comprises 87 and OpSim\_ roll.cad\_0.8\_r\_RA\_0.0\_D\_-10.0  204 observations over 10 yr. The mean sampling of  u-band cadences, such as $baseline\_u\_ra\_0.0\_de\_-10.0,baseline\_u\_ra\_0.0\_de\_-30.0, rolling\_cadence\_0.8\_u\_ra\_0.0\_de\_-10.0., rolling\_cadence\_0.8\_u\_ra\_0.0\_de\_-30.0$, are 58.2, 75.1, 64.8, 66.7 days respectively. Such condition limit the interpretation of the AGN time-lag and periodicity analysis.
 Our metrics indicate that denser rolling cadences are more reliable for extraction of finer level of information such as AGN variability observables.}
 \vspace{-0.35cm}
\begin{quote}
{\footnotesize Q7:  Are there any science drivers pushing for or against particular dithering patterns (either rotational dithers or translational dithers?)  If available, please mention specific simulated cadences, and specific metrics, that support your answer.} 
\end{quote}
\vspace{-0.35cm}
{ Answer 7 
Not applicable.}
\bibliography{sdsstech,richards_longnames,richards_shortnames,refs}{}
\bibliographystyle{aasjournal}

{\bf [Remove this notice after trimming to 3 pages.]}

\end{document}